\documentclass[prb,aps,twocolumn,showpacs]{revtex4}
\usepackage{amsmath,amssymb}
\usepackage{graphicx}
\begin{document}

\title{Remarkable magnetostructural coupling around the magnetic transition in CeCo$_{0.85}$Fe$_{0.15}$Si}

\author{V. F. Correa}
\author{D. Betancourth}
\author{J. G. Sereni}

\affiliation{Centro At\'omico Bariloche (CNEA) and Instituto Balseiro (U. N. Cuyo), 8400 Bariloche, R\'io Negro, Argentina}

\author{N. Caroca Canales}
\author{C. Geibel}

\affiliation{Max-Planck-Institut f\"ur Chemische Physik fester Stoffe, D-01187 Dresde, Germany}

\date{\today}


\begin{abstract}

We report a detailed study of the magnetic properties of CeCo$_{0.85}$Fe$_{0.15}$Si under high magnetic fields (up to 16 Tesla) measuring different physical properties such as specific heat, magnetization, electrical resistivity, thermal expansion and magnetostriction.
CeCo$_{0.85}$Fe$_{0.15}$Si becomes antiferromagnetic at $T_N \approx$ 6.7 K. However, a broad tail (onset at $T_X \approx$ 13 K) in the specific heat precedes that second order transition. This tail is also observed in the temperature derivative of the resistivity. However, it is particularly noticeable in the thermal expansion coefficient where it takes the form of a large bump centered at $T_X$.
A high magnetic field practically washes out that tail in the resistivity. But surprisingly, the bump in the thermal expansion becomes a well pronounced peak fully split from the magnetic transition at $T_N$. 
Concurrently, the magnetoresistance also switches from negative to positive just below $T_X$. 
The magnetostriction is considerable and irreversible at low temperature ($\frac {\Delta L}{L} \left(16 T\right) \sim$ 4$\times$10$^{-4}$ at 2 K) when the magnetic interactions dominate. A broad jump in the field dependence of the magnetostriction observed at low $T$ may be the signature of a weak ongoing metamagnetic transition.  
Taking altogether the results indicate the importance of the lattice effects on the development of the magnetic order in these alloys.    
    
\end{abstract}

\maketitle

\section{introduction}

The wide diversity of ground states observed in intermetallic Cerium-based compounds arises mostly as a consequence of the dual localized-itinerant character of the electrons in the partially filled 4$f$ orbitals.
The interaction between 4$f$ electrons and conduction electrons from other orbitals (s,p,d) often ends up in two different and contrasting behaviors:$\cite{Fazekas}$ (i) a magnetic ordered state of well localized magnetic moments , or (ii) a non-magnetic ``singlet'' state.
The former realizes when the so-called Ruderman-Kittel-Kasuya-Yosida interaction is dominant. 
The ``singlet'' state, on the other hand, is a sort of generalized Kondo effect (the Kondo lattice) arising from the hybridization with the conduction electrons which usually gives raise to large effective masses.



CeCo$_{1-x}$Fe$_{x}$Si alloys are a good example where both extreme behaviors, i.e. the Kondo lattice-heavy fermion and the magnetic order, can be observed upon changes in the relative concentration between Co and Fe.\cite{Sereni2014} While CeCoSi shows a textbook second order antiferromagnetic transition ($T_N =$ 8.8 K) with well localized electrons in the 4$f$ orbitals of trivalent Ce$^{3+}$ ions,\cite{Chevalier2004} CeFeSi is non-magnetic\cite{Welter1992} and, despite there are no specific heat reported measurements, it is expected to be a moderate heavy fermion with a Sommerfeld coefficient $\gamma \sim$ 100 mJ/mol$\cdot$K$^2$.\cite{Sereni2014}  

\begin{figure}[!h]
\includegraphics[width=\columnwidth]{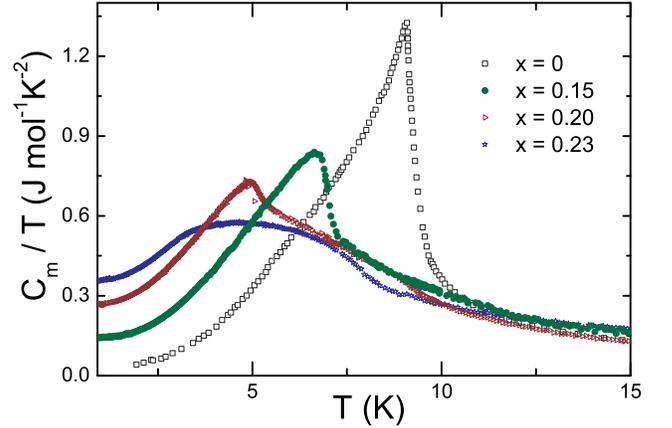}
\caption[]{(color online) Magnetic contribution to the specific heat divided by temperature at selected Fe doping levels $x$ (adapted from Ref. [\onlinecite{Sereni2014}]). The lattice vibrations contribution has been substracted from the isotypic La compounds.}
\label{fig1}
\end{figure}

Increasing the Fe content $x$ gradually suppresses the magnetic order. Nonetheless, at intermediate concentrations still on the Co-rich side, a broad magnetic tail is seen in the specific heat just above $T_N$ (see Fig. \ref{fig1} for $x$ = 0.15). The tail is reminiscent of short-range correlations and it was ascribed to a dimensional crossover of the magnetic fluctuations.\cite{Sereni2014}
Further increasing $x$ shows a full development of this tail into a broad bump anomaly (see Fig. \ref{fig1} for $x$ = 0.23) concomitant with the complete suppression of the magnetic order. Above $x \approx$ 0.35 even the anomaly disappears and the low temperature specific heat shows strong non-Fermi liquid behavior. Finally, a heavy Fermi-liquid is recovered when $x$ approaches 1. \cite{Sereni2014}

In this work we focus on the effects of a high magnetic field on the magnetic and lattice properties of CeCo$_{0.85}$Fe$_{0.15}$Si using different experimental techniques such as specific heat, magnetization, electrical resistivity, thermal expansion and magnetostriction. This selected concentration is particularly interesting because both the magnetic order and the broad tail are observed. 
Even though these magnetic features are observed in all the experiments, it is the atomic lattice that is markedly sensitive to the magnetic tail above $T_N$, as seen in the thermal expansion measurements.
 
The applied magnetic field has dissimilar effects on the magnetic order and the broad tail. While the former hardly feels the magnetic field, the tail anomaly is deeply affected by fields larger than 10 Tesla. 
Moreover, the effect varies along the different experiments. Whilst the tail in the resistivity vanishes in a magnetic field, it evolves to a distinct well pronounced peak in the thermal expansion coefficient. This feature is suggestive of a phase transition and shows the large influence of the lattice in the formation of the ground state of these alloys. 
In this direction, a broad and smooth jump observed in the field dependence of the low temperature magnetostriction may be the signature of a weak ongoing metamagnetic transition.  

\section{experimental details}

Polycrystalline samples of CeCo$_{0.85}$Fe$_{0.15}$Si were prepared by arc melting stoichiometric amounts of the pure elements followed by an annealing procedure as it is described elsewhere.\cite{Sereni2014}
A standard heat-pulse technique was used in the specific heat experiments. Magnetization measurements were carried out in a commercial Quantum Design PPMS magnetometer. A high resolution capacitive dilatometer was used in the dilation experiments while a stantard four probe technique was used in the  electric transport measurements.
Magnetoresistance and magnetostriction were measured in a 18 Tesla-superconducting magnet down to 1.5 K. 
A bar-shaped sample was cut for the electrical resistivity experiments while a sample with cubic-like shape was used in the dilatometry experiments.
All the dilation experiments under a magnetic field where carried out in the longitudinal configuration, i.e. with the magnetic field $B$ parallel to the sample dimension $L$ being measured.

\section{results}

\begin{figure}[t]
\includegraphics[width=\columnwidth]{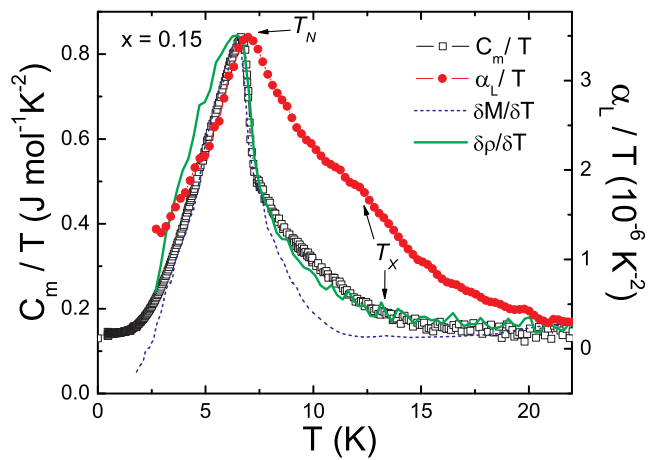}
\caption[]{(color online) Temperature dependence of the magnetic specific heat divided by temperature (left axis) and the linear thermal expansion coefficient divided by temperature (right axis) for $x=$ 0.15. The temperature derivatives of the magnetization and the resistivity in arbitrary units are also shown.}
\label{fig2}
\end{figure}

Figure \ref{fig2} displays a comparison of the observed low temperature properties of CeCo$_{0.85}$Fe$_{0.15}$Si between different experiments: magnetic specific heat $C_m$, linear thermal-expansion coefficient $\alpha_L = \frac{1}{L}\left(\frac {\partial \Delta L}{\partial T} \right)$, and temperature derivatives of magnetization $M$ and electrical resistivity $\rho$. 
The magnetic transition at $T_N \approx$ 6.7 K is clearly detected by all experiments. The onset of the broad tail in $C_{m}/T$ occurs at $T_X \approx$ 13 K. The temperature derivative of $\rho$ basically traces $C_{m}/T$. On the other hand, though the magnetic transition is well identified in $\partial M/\partial T$, the tail is hardly seen.
But notably, the atomic lattice is especially sensitive to whatever is responsible for the tail-anomaly. This can be inferred by the major effect it has on the thermal expansion coefficient. There is even a well pronounced kink in $\alpha_L / T$ around $T_X$.

\begin{figure}[b]
\includegraphics[width=\columnwidth]{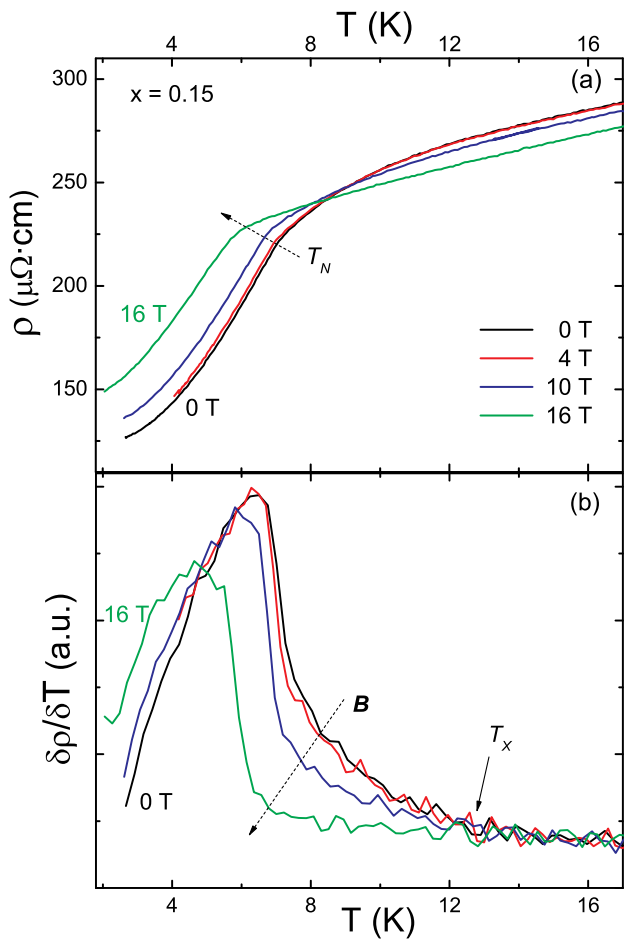}
\caption[]{(color online) Temperature dependence of the resistivity (a) and its derivative (b), at different applied magnetic fields.}
\label{fig3}
\end{figure}

To get further insight into the nature of the tail anomaly we study the response of CeCo$_{0.85}$Fe$_{0.15}$Si to high magnetic fields. The temperature dependence of the electrical resistivity and its derivative at different fields are displayed in Figs. \ref{fig3} (a) and (b), respectively. The large residual resistivities observed are not intrinsic, but related to the high proliferation of micro-cracks in the polycrystalline samples. 

\begin{figure}[t]
\includegraphics[width=\columnwidth]{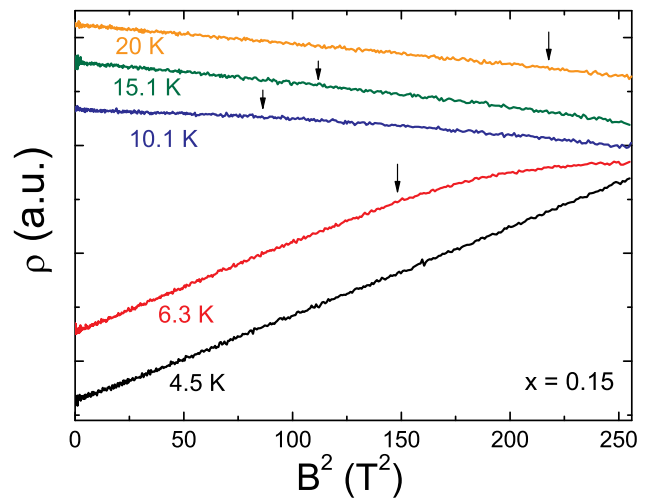}
\caption[]{(color online) Magnetoresistivity versus square magnetic field at different temperatures. Arrows indicate the extent of the linear dependence. Curves have been vertically shifted.}
\label{fig4}
\end{figure}

There are several findings worth mentioning. 
First, a magnetic field $B \lesssim$ 5 T (not shown in Fig. \ref{fig3}) has no effect on the transport properties. This is in agreement with previous specific heat experiments under magnetic fields.\cite{Sereni2014} 
Second, the robustness of the magnetic order: in 16 Tesla, the ordering temperature decreases just 2 K (see Fig. \ref{fig3} (b)). 
Third, and more interesting, is the fact that between $T_N$ and $T_X$ the magnetorresistance switches from negative (at higher $T$) to positive (at lower $T$). 
Figure \ref{fig4} shows that the magnetorresistance has a quadratic dependence with $B$ at temperatures far away enough from the inversion temperature, but it ceases to have this dependence around that temperature.
 
The most remarkable finding in Fig. \ref{fig3} (b), however, is the gradual reduction of the tail anomaly as the field is increased until it fully  disappears at $B \lesssim$ 16 T. 

\begin{figure}[t]
\includegraphics[width=\columnwidth]{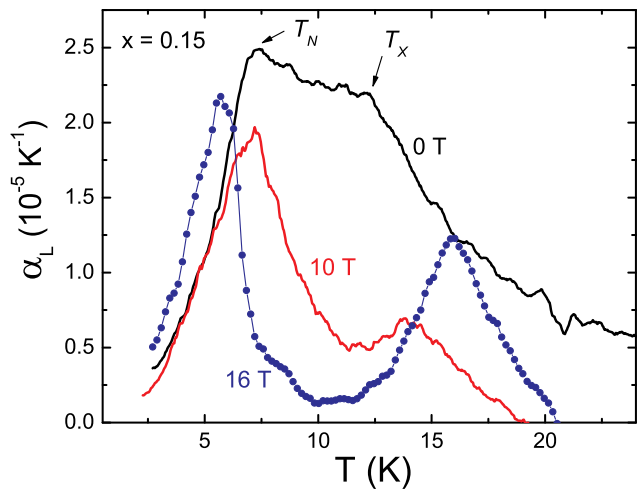}
\caption[]{(color online) Linear thermal expansion coefficient versus temperature at different applied magnetic fields in the longitudinal configuration.}
\label{fig5}
\end{figure}

The question arises about what happens with the atomic lattice. Surprisingly, the magnetic field imprint on the lattice shows little correlation to what happens to the electrical resistivity, mostly above $T_N$. Figure \ref{fig5} shows the linear thermal expansion coefficient at different applied magnetic fields.     
Again, it is evident that the magnetic order (characterized by $T_N$) is hardly affected by the magnetic field (in fact, though not shown in Fig. \ref{fig5}, $\alpha_L$ is completely insensitive to magnetic fields $B \lesssim$ 5 T). 
On the other hand, the broad tail (characterized by $T_X$) is highly reduced in an intermediate magnetic field of 10 T. At higher fields, however, the broad tail re-emerges, now as a very distintictive peak in $\alpha_L$ (suggestive of a phase transition) which is well split from the magnetic order peak as seen in Fig. \ref{fig5} (see the curve at $B =$ 16 T). 
This result is astonishing since absolutely no trace of an eventual transition is seen in the resistivity. 
Other high field thermodynamic measurement like magnetization or specific heat would be highly valuable to further check the possibility of a field induced transition. Yet, the magnetoelastic nature of the tail anomaly is evident.  

\begin{figure}[b]
\includegraphics[width=\columnwidth]{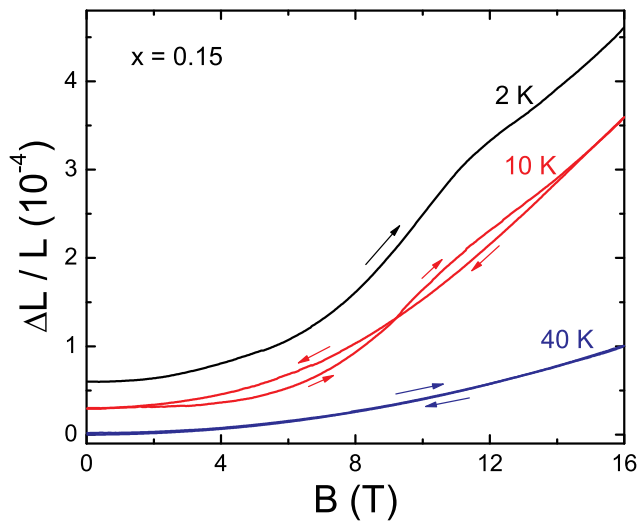}
\caption[]{(color online) Linear magnetostriction versus magnetic field at different temperatures in the longitudinal configuration. Arrows indicate the direction of the field sweep. Curves have been vertically shifted.}
\label{fig6}
\end{figure}

Further insight in this direction can be obtained from magnetostriction experiments. Figure \ref{fig6} displays the longitudinal magnetostricition at different fixed temperatures. The magnitude of the magnetostrictive effect below 10 K ($\frac {\Delta L}{L} \left(16 T\right) \sim$ 4$\times$10$^{-4}$ at 2 K) is quite important even compared to other Ce-compounds.
At high enough temperature, where the magnetic correlations vanish, the magnetostriction has a quadratic field dependence (see the curve at 40 K). At lower $T$ ($\sim T_X$), the magnetostriction deviates from that dependence, becomes hysteretic and starts to develop a broad and subtle kink around 10 Tesla in the up-sweep. This kink, even still smooth, becomes more evident at lower temperature (see the curve at 2 K) and it is reminiscent of a metamagnetic transition.\cite{metamagnetism}

\section{discussion}

The nature of the tail-anomaly above $T_N$ is debatable. But its overall bump-like shape in the specific heat, along with its rapid suppression in a magnetic field (as seen in $\partial \rho / \partial T$) suggest that it may be related to short-range magnetic correlations, probably quasi-two-dimensional as can be speculated from the crystal structure.\cite{Sereni2014} 

A similar bump anomaly has already been observed in electrical resistivity experiments performed on stoichiometric CeCoSi polycrystals under pressure.\cite{Lengyel2013} Even though there is a huge difference in the residual resistivity ratio between those non-doped samples ($\sim$ 170) and our alloys ($\sim$ 3), the response to an applied magnetic field shows interesting similarities: the tail anomaly in $\partial \rho / \partial T$, for instance, gets completely washed out at 9 T in CeCoSi. Unlike our results, however, the magnetic order (identified as a peak in $\partial \rho / \partial T$ at $T_N$) becomes much more pronounced in a magnetic field.   
This observation effectively suggests that in CeCoSi the same electron reservoir is involved in both the magnetic order and the tail anomaly. This is not clear in our case.
Moreover, the magnetoresistance in CeCoSi is positive below $T_X$ and becomes negligible at higher temperature.\cite{Lengyel2013} This high-$T$ behavior contrasts with what is observed in our alloy: a clear switch from a positive to a negative magnetoresistance around $T_X$ which may be related to the presence of competing types of magnetic couplings.

Nonetheless, the most relevant result of this work is the notably large coupling of the tail-anomaly to the atomic lattice and its peculiar evolution with a magnetic field. As seen in the thermal expansion coefficient (Fig. \ref{fig5}), the anomaly weakens in a moderate magnetic field ($B \lesssim$ 10 T). But, at higher fields, it clearly invigorates and evolves into a well defined peak suggesting a magnetically-driven lattice distortion.

Large magnetostructural effects have already been observed in a related compound belonging to the same 111 family and crystallizing in the same crystal structure: CeTiGe. It is a non-magnetic heavy-fermion system that presents a first order-like metamagnetic transition strongly coupled to the atomic lattice ($\Delta L / L \sim$ 3$\times$ 10$^{-3}$).\cite{Deppe2012}
In this sense, that observation also supports the idea of a weak incipient metamagnetism in our magnetostriction results. If in a ``simpler'' system (non-magnetically ordered, i.e., higher Fe concentration) this phenomenon evolves into a very distinctive metamagnetic phase transition and a larger magnetostructural coupling as in CeTiGe, is a matter of ongoing study.

\section{conclusions}

In summary, we have presented an experimental study of the magnetic properties of CeCo$_{0.85}$Fe$_{0.15}$Si using different experimental techniques. The antiferromagnetic order ($T_N \approx$ 6.7 K) is preceded by a tail detected in the specific heat and the temperature derivative of the resistivity. But, it is in the thermal expansion coefficient where it appears as a particularly large anomaly above $T_N$.
An applied magnetic field has dissimilar effects on the magnetic order and the tail. While the former hardly feels the magnetic field, the tail anomaly is deeply affected by fields larger than 10 Tesla. 
Moreover, the effect varies along the different experiments. Whilst the tail in the resistivity vanishes in a magnetic field, it evolves to a distinct well pronounced peak in the thermal expansion coefficient (with a sizable magnetostrictive magnitude), which may be the signature of a structural distortion driven by magnetic correlations. 
At the same time, a broad and smooth jump observed in the field dependence of the low temperature magnetostriction may indicate a weak ongoing metamagnetic transition.
Altogether, these observations confirm that the atomic lattice plays a prominent role in the formation of the ground state of these alloys. 

\section{Acknowledgments}

Authors thank D. J. Garc\'ia and P. S. Cornaglia for helpful discussions. V. F. C and J. G. S. are members of CONICET, Argentina. Work partially supported by ANPCyT PICT2010-1060 Bicentenario, SeCTyP-UNCuyo 06/C457 and C002.

\end{document}